\documentclass[letterpaper,10pt]{article} 

\usepackage{opticameet3} 

\newcommand\authormark[1]{\textsuperscript{#1}}

\usepackage{amsmath,amssymb}
\usepackage{subcaption}
\usepackage{makecell}
\usepackage{soul}
\usepackage[colorlinks=true,bookmarks=false,citecolor=blue,urlcolor=blue]{hyperref} 

\begin{document}

\title{Adaptive Turbo Equalization of Probabilistically Shaped Constellations}

\vspace{-0.5cm}
\author{Edson Porto da Silva \authormark{1,*}, Metodi Plamenov Yankov \authormark{2}}

\address{\authormark{1} Federal University of Campina Grande (UFCG), Campina Grande, 58429-900, Brazil.\\
\authormark{2}Technical University of Denmark, DTU Electro, Kgs. Lyngby, 2800, Denmark.}

\email{\authormark{*}edson.silva@dee.ufcg.edu.br} 
\vspace{-0.8cm}
\begin{abstract}
Fiber nonlinearity compensation of probabilistically shaped constellations with adaptive turbo equalization is investigated for the first time. Potential for more than 100\% transmission reach extension is demonstrated by combining probabilistic shaping, single-channel digital backpropagation, and adaptive turbo equalization.
\end{abstract}
\vspace{-0.2cm}
\section{Introduction}
Fiber nonlinearity compensation (NLC) is a key enabler for reach and data rate maximization of coherent wavelength division multiplexing (WDM) systems over single-mode fibers. Nevertheless, despite the progress made over the years, finding ways of applying NLC effectively to overcome fiber nonlinear interference (NLI) is still a work in progress, with most solutions targeting the intra-channel NLI \cite{Cartledge2017}. Nowadays, state-of-the-art coherent optical transceivers combine probabilistically shaped (PS) constellations and powerful forward error correction (FEC) to maximize transmission throughput and reach \cite{InfineraWP}. In this scenario, the joint design of modulation, FEC and NLC is necessary to achieve higher transmission performance. Recently, turbo equalization schemes (TEq) have been proposed for this purpose \cite{Golani2019,Silva2021}, achieving compensation of inter-channel NLI in WDM systems. In a TEq scheme, the receiver iterates between channel equalization and FEC decoding, aiming to improve the performance of both functionalities by exchanging soft-information from decoder to equalizer and vice-versa. It has been demonstrated that TEq has the potential to increase throughput \cite{Golani2019} and transmission reach \cite{Silva2021} in coherent WDM systems using uniform quadrature and amplitude (QAM) constellations. However, the use of TEq has not been yet investigated for PS QAM constellations.

In this paper, the TEq scheme from \cite{Silva2021} is analyzed for PS signals, and its effectiveness demonstrated to be higher than for uniform signals. The combined performance of PS, digital backpropagation (DBP), and adaptive TEq is evaluated, highlighting the potential for transmission reach extension of WDM systems.
\vspace{-0.2cm}
\section{Turbo equalization method}
\begin{figure}[b]
\vspace{-0.4cm}
    \centering
    \includegraphics[width=0.85\linewidth]{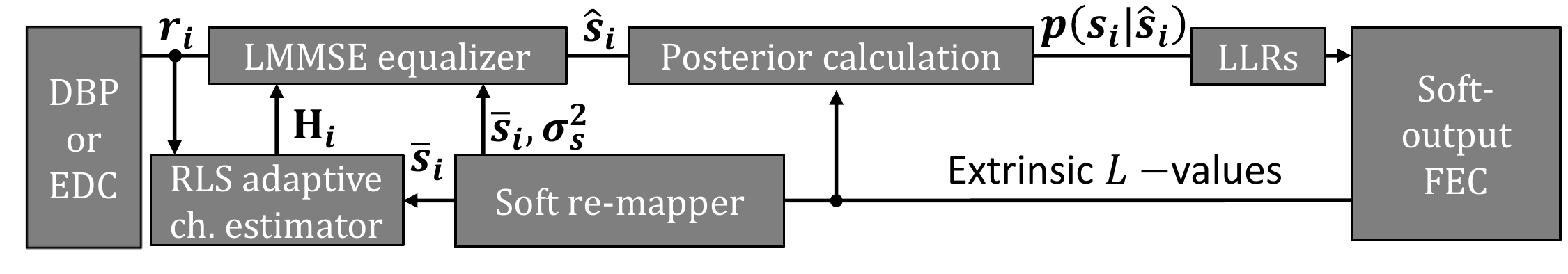}
    \caption{Block diagram of the turbo equalizer used in this work.}
    \label{fig:BD}
\end{figure}

\vspace{-0.1cm}
The high-level description of the employed turbo equalizer is given in Fig.~\ref{fig:BD}. After single shot DBP or electronic dispersion compensation (EDC), the assumed channel model takes the shape $\mathbf{r}_i = \mathbf{H}_i \mathbf{s}_i + \mathbf{n}_i$, where $i$ represents time instant, and $\mathbf{r}_i, \mathbf{s}_i$ and $\mathbf{n}_i$ are vectorized, dual polarization received symbols, transmitted symbols and channel noise, respectively. The sizes of these vectors are $2N$, where $N$ is the length of the 1 sample per symbol equalizer. The matrix $\mathbf{H}$ is the assumed linear interference matrix of an appropriate size. 
The extrinsic log-likelihood ratios (LLRs) out of the FEC decoder are used to estimate soft-symbols, represented by mean $\bar{s}$ and variance $\sigma^2_s$. This can be thought of as a soft mapping function. A recursive least squares (RLS) algorithm is used to estimate channel coefficients $\mathbf{H}_i$ at time $i$, assuming a linear interference channel model $\mathbf{r}_i = \mathbf{H}_i \bar{\mathbf{s}}_i$. The estimated channel coefficients are then used to obtain a linear minimum mean squared error (LMMSE) equalizer coefficients $\mathbf{w}_{i}$, which are then used to obtain equalized symbols $\hat{s}_{i}=\mathbf{w}_{i}^{H}\left[\mathbf{r}_{i}-\mathbf{H}_i \overline{\mathbf{s}}_{i}\right]$. 

In order to obtain posterior symbol distributions, an equivalent channel model is assumed of the form $\mathbf{\hat{s}}_i = \mathbf{M}_i \mathbf{s}_i + \mathbf{\eta}_i$. The noise term $\mathbf{\eta}_i$ is assumed Gaussian, leading to a Gaussian likelihood function $p(\mathbf{\hat{s}}_i|\mathbf{s}_i)$. When probabilistic shaping is applied, the posterior distirbution of the equivalen channel is obtained as $p(\mathbf{s}_i|\mathbf{\hat{s}}_i) \propto p(\mathbf{\hat{s}}_i|\mathbf{s}_i)p(\mathbf{s}_i)$, where $p(\mathbf{s}_i)$ is the prior distribution of the symbols. The posterior distributions are then used to obtain LLRs for the next FEC iteration using conventional bit-metric decoding. For space considerations, further details on the derivation of the LMMSE coefficients and the parameters of this equivalent model (i.e. coefficients of the equivalent matrix $\mathbf{M}$ and covariance matrix of the noise term $\mathbf{\eta}$) are omitted, and can be found here \cite{Silva2021}.
\vspace{-0.2cm}
\section{Simulation setup and results}
\vspace{-0.1cm}
The simulation setup assumes a transmission of eleven polarization-multiplexed (PM) WDM carriers modulated at $\mathrm{32~GBd}$ in a $\mathrm{37.5~GHz}$ frequency grid spacing. The sequence of shaped constellation symbols is generated by encoding pseudo random bit sequences with a constant composition distribution matcher (CCDM) \cite{Schulte2016}. The soft-FEC is implemented with an ARJ4A low-density parity-check (LDPC) code \cite{Divsalar2005}. The range of FEC code rates used is obtained by using different puncturing rates. The coded symbols in each LDPC frame are interleaved before upsampling and pulse shaping. Pilot-symbols for equalization and carrier phase recovery are uniformly inserted in between coded symbols at a rate of $5\%$. An upsampling rate of $\mathrm{16~samples/symbol}$ and pulse shaping with a root-raised cosine (RRC) filter with a roll-off factor of $\mathrm{0.01}$ are applied. Every Monte Carlo trial runs a sequence of symbols consisting of $\mathrm{18}$ LDPC code blocks per polarization. Performance metrics shown in the results are averaged over five Monte Carlo trials ($>\mathrm{2.2 M}$ information bits used in total). The nonlinear fiber channel is simulated using the symmetric split-step Fourier (SSFM) to solve the Manakov fiber model \cite{Marcuse1997}. The model assumes transmission over multiple $\mathrm{50~km}$ spans of standard, single mode fiber, with fiber attenuation compensated by Erbium-doped fiber amplifiers (EDFAs) with a noise figure of $\mathrm{4.5~dB}$. The SSFM is set to run with a fixed $\mathrm{100~m}$ step-size. The fiber parameters of attenuation, nonlinear coefficient and chromatic dispersion are chosen to be $\mathrm{\alpha}=~0.2~\mathrm{dB/km}$, $\mathrm{\gamma}=~1.~3~\mathrm{W^{-1}km^{-1}}$, and $\mathrm{D}=~17~\mathrm{ps/nm/km}$, respectively. Local oscillators are assumed to be ideal and dynamic polarization effects were not considered. At the receiver, the performance of the central WDM channel is evaluated. The same sequence of digital signal processing (DSP) blocks used in \cite{Silva2021} is assumed at the receiver, adapted to process PS signals. Moreover, the bit interleaver assumed in \cite{Silva2021} is replaced by a symbol interleaver instead, to make the modulation scheme compatible with CCDM. To compare performance of uniform and shaped constellations, the net information rate (IR) is fixed is $\mathrm{6.5~bits/QAMsymbol}$ and different configurations of FEC rate and constellation entropy are set, as depicted in Table~\ref{tab:Tab1}.

\begin{table}[h!]
    \vspace{-0.2cm}
    \centering
    \resizebox{0.65\textwidth}{!}{
    \begin{tabular}{|c|c|c|c|c|c|}
    \hline Modulation & Entropy [bits] & $\mathrm{k}$ & $\mathrm{n}$ & $\mathrm{R}=\mathrm{k} / \mathrm{n}$ &  IR [bits/symb] \\
    \hline shaped 1024QAM & 8.5 & 16384 & 20480 & 0.8000 & 6.50 \\
    \hline uniform 256QAM & 8 & 16384 & 20160 & 0.8127 & 6.50 \\
    \hline uniform 1024QAM & 10 & 16384 & 25200 & 0.6502 & 6.50 \\
    \hline
    \end{tabular}}
    \vspace{-0.1cm}
    \caption{Modulation and FEC parameters of the three tested configurations.}
    \label{tab:Tab1}
    \vspace{-0.7cm}
\end{table}

\begin{figure*}[b!]
    \vspace{-0.5cm}
    \centering
    \begin{subfigure}[h]{0.5\textwidth}
        \centering
        \includegraphics[width=\linewidth]{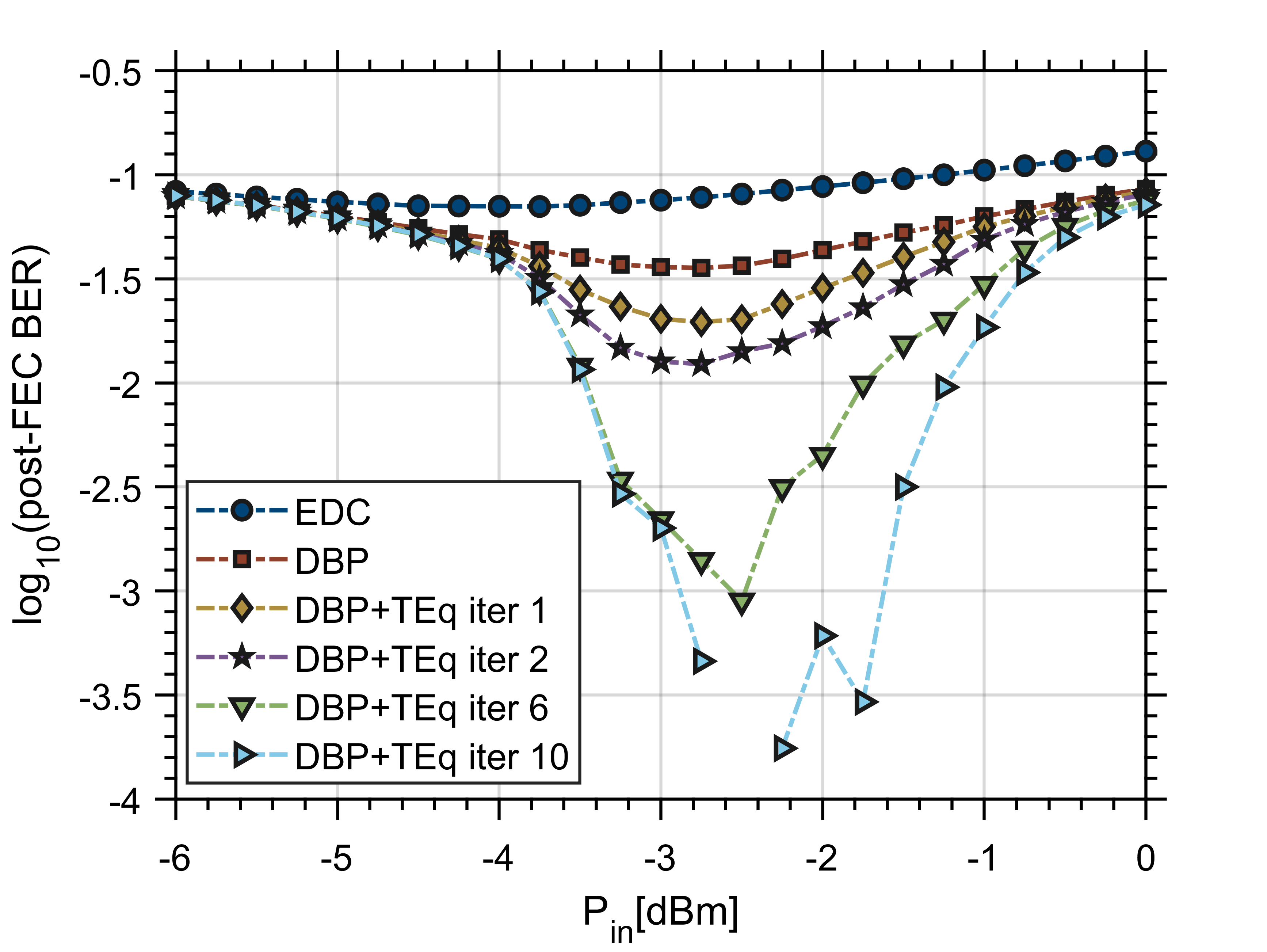}
        \caption{post-FEC BER versus fiber input power.}
        \label{fig:BERvsPin}
    \end{subfigure}%
    ~ 
    \begin{subfigure}[h]{0.5\textwidth}
        \centering
        \includegraphics[width=\linewidth]{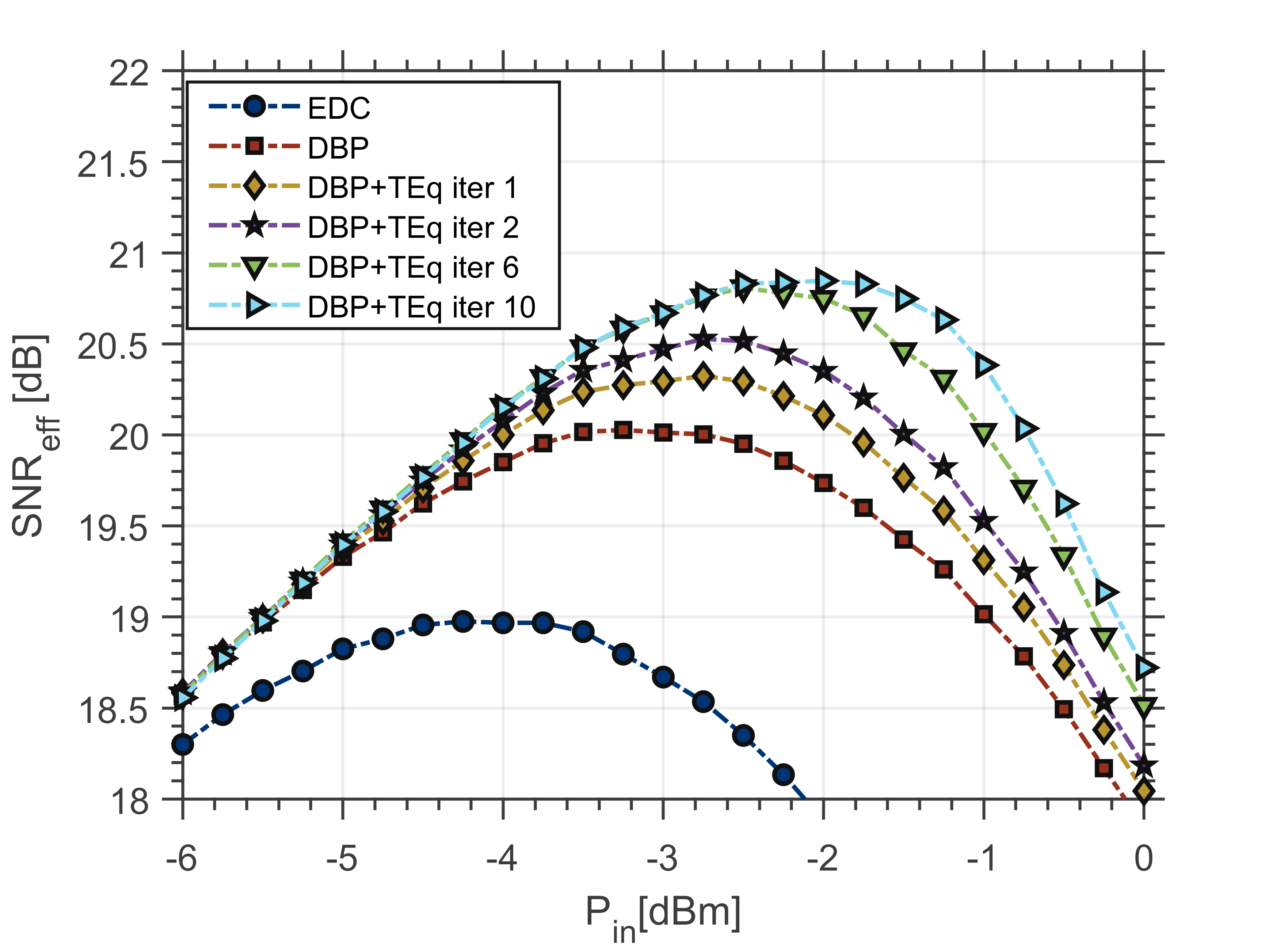}
        \caption{$\mathrm{SNR_{eff}}$ versus fiber input power.}
        \label{fig:SNRvsPin}
    \end{subfigure}
    \vspace{-0.6cm}
    \caption{Central WDM channel performance of the PS PM-1024QAM signals after 29$\times$50~km of transmission.}
    \label{fig:Fig3}
\end{figure*}

Shaped 256QAM is not included in the analysis because, to achieve the target net IR, the required FEC code rate beyond the viable rates achieved by puncturing the LDPC code assumed in this work. Nevertheless, for the targeted net IR, PS 1024QAM would be the best choice in terms of flexibility and rate adaptation. Moreover, uniform 1024QAM is included as basis for comparison of the PS 1024QAM performance.

\begin{figure}[h!]
    \centering
    \vspace{-0.2cm}
    \includegraphics[width=1\linewidth]{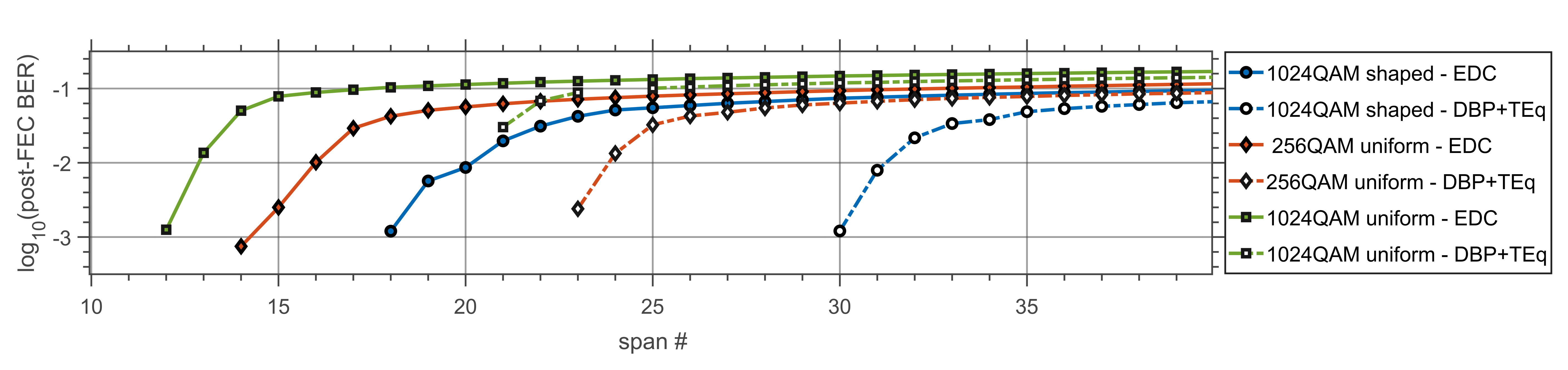}
    \vspace{-0.9cm}
    \caption{Post-FEC BER as a function of the number of propagated fiber spans for all three investigated configurations.}
    \label{fig:Fig4}
    \vspace{-0.5cm}
\end{figure}

In Fig.~\ref{fig:BERvsPin} and Fig.~\ref{fig:SNRvsPin}, performance curves of BER and effectively received signal-to-noise ratio ($\mathrm{SNR_{eff}}$) as a function of fiber launch power per WDM channel are shown for the PS PM-1024QAM signals after 29$\times$50~km of transmission. Here, $\mathrm{SNR_{eff}}$ is the average SNR estimated at the output of the LMMSE equalizer. After 10 TEq iterations, the $\mathrm{SNR_{eff}}$ is improved by extra $\approx \mathrm{0.80~dB}$ on top of $\mathrm{1.0~dB}$ of the DBP gain, leading to BER = 0 for power values close to $\mathrm{-2.0~dBm}$. In Fig.~\ref{fig:Fig4}, the transmission reach extension of receivers employing DBP+TEq is compared to receivers that only implement EDC. For the uniform constellations, the transmission reach is extended by $\mathrm{8~spans~(400~km)}$, whereas $\mathrm{11~spans~(550~km)}$ are gained for PS 1024QAM. The extra gain for the PS signals could be related to the fact that shaped constellations tend to enhance the NLI in the nonlinear fiber channel. Overall, at the same net IR, the combined gains of PS, single-channel DBP, and TEq have allowed $\mathrm{(29-13)/13}\approx 120\%$ of transmission reach extension, compared to EDC. It should be noted that all results presented are constrained to at most 10 TEq iterations. However, although most of the performance improvement is achieved within this number of iterations, for operations around the optimum launch power and in the highly nonlinear regime, marginal improvements in BER are still possible with more iterations performed.

Next, the benefits of compensating the inter-channel NLI with TEq are evaluated. In Table~\ref{tab:Tab2}, the $\mathrm{SNR_{eff}}$ at the decoding threshold are presented for all modulation schemes detailed in Table~\ref{tab:Tab1}. The decoding threshold is defined as the minimum $\mathrm{SNR_{eff}}$ necessary to obtain post-FEC $\mathrm{BER=0}$ in the simulations. From the results in Table~\ref{tab:Tab2}, it can be noted that after the nonlinear fiber channel, if the receiver is set to perform EDC only, all modulations show a penalty from $\mathrm{0.2}$ to $\mathrm{0.5~dB}$ w.r.t the AWGN channel performance. By replacing EDC with DBP, the $\mathrm{SNR_{eff}}$ increases, but the decoding threshold also goes up. This effect is most probably due to the fact that, after compensating for intra-channel NLI, the remaining inter-channel NLI would appear as time-varying intersymbol interference (ISI), increasing the mismatch between the true channel and the memoryless Gaussian auxiliary channel model assumed by the soft-decoder \cite{Borujeny2021}. Thus, the increase in $\mathrm{SNR_{eff}}$ after DBP does not translate linearly to post-FEC BER improvement. Finally, if DBP is followed by TEq, the decoding threshold is completely recovered for 256QAM and almost completely recovered for shaped 1024QAM ($\mathrm{0.2~dB}$ penalty remains). Hence, by reducing the impact of the time-varying ISI, the TEq also helps the mismatched soft-decoding process at the receiver.

\begin{table}[h!]
    \vspace{-0.2cm}
    \centering
    \resizebox{0.9\textwidth}{!}{
    \begin{tabular}{|c|c|c|c|c|}
        \hline Modulation & \makecell{SNR @ dec. threshold \\ (AWGN ch.)} & \makecell{$\mathrm{SNR_{eff}}$ @ dec. threshold \\  (fiber ch. $+\mathrm{EDC}$ )} & \makecell{$\mathrm{SNR_{eff}}$ @ dec. threshold \\ (fiber ch. $+\mathrm{DBP})$} &\makecell{$\mathrm{SNR_{eff}}$ @ dec. threshold \\  (fiber ch. $+\mathrm{DBP}+$ TEq)}\\
        \hline shaped 1024QAM & $\approx 20.7 \mathrm{~dB}$ & $\approx21.2 \mathrm{~dB}$ &    $\approx 22.1 \mathrm{~dB}$ & $\approx 20.9 \mathrm{~dB}$ \\
        \hline uniform 256QAM & $\approx 21.9 \mathrm{~dB}$ & $\approx 22.4 \mathrm{~dB}$ &     $\approx 22.8 \mathrm{~dB}$ & $\approx 21.9 \mathrm{~dB}$ \\
        \hline uniform 1024QAM & $\approx 22.8 \mathrm{~dB}$ & $\approx 23.0 \mathrm{~dB}$ & $\approx 23.6 \mathrm{~dB}$ & $\approx 22.6\mathrm{~dB}$\\
        \hline
    \end{tabular}}
    \vspace{-0.2cm}
    \caption{Required SNR at the decoding threshold for different modulations schemes and receiver configurations.}
    \label{tab:Tab2}
    \vspace{-0.9cm}
\end{table}
\vspace{-0.2cm}
\section{Conclusions}

The effectiveness of adaptive turbo equalization for nonlinearity compensation was assessed for probabilistically shaped constellations. It is shown that. Potential for more than $\mathrm{100}\%$ extension in transmission reach is verified in simulations by combining probabilistic shaping, digital backpropagation and adaptive turbo equalization.
\vspace{-0.2cm}
\bibliographystyle{IEEEtran}
\bibliography{Bibliography}

\end{document}